\documentclass[aps,prl,twocolumn,superscriptaddress]{revtex4}
\usepackage[]{graphicx}

\begin{document}

\begin{widetext}
\begin{figure*}
\includegraphics[width=17.4cm]{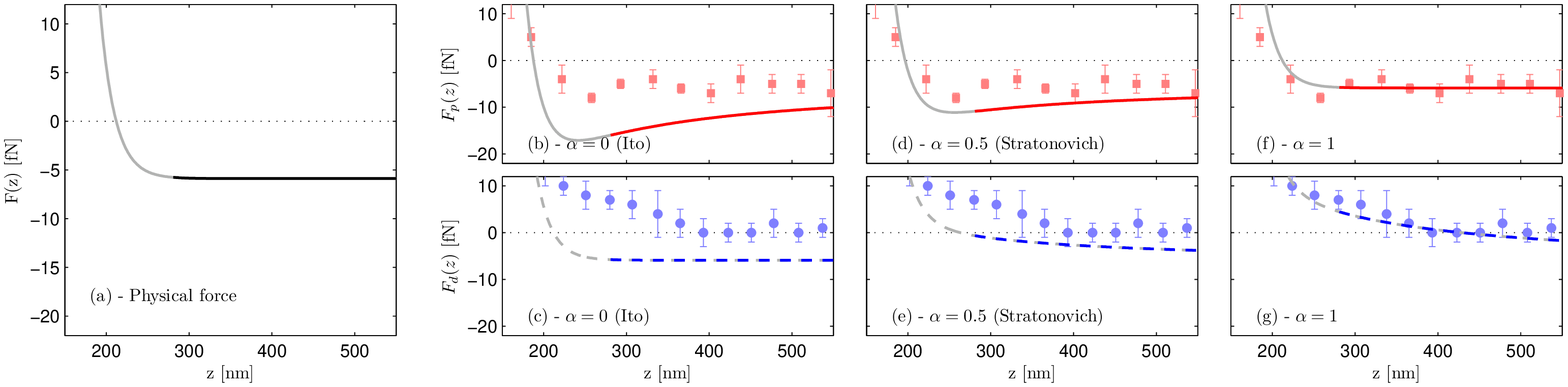}
\caption{(color online) (a) Force acting on the Brownian particle studied in Ref. \cite{Volpe2010}; the black line highlights the region where only gravity (buoyancy) acts. (b-g) Expected $F_p(z)$ (solid red line) and $F_d(z)$ (dashed blue line) for $\alpha$ = 0, 0.5 and 1; the symbols represent the experimental measurement of $F_p(z)$ (red squares) and $F_d(z)$ (blue circles) (from Fig. 2 in Ref. \cite{Volpe2010}).}
\end{figure*}
\end{widetext}

\textbf{Volpe \emph{et al.} Reply:} In our Letter \cite{Volpe2010} we have studied a Brownian particle diffusing in front of a horizontal wall, whose movement $z(t)$ is described by a stochastic differential equation (SDE) with multiplicative noise. This SDE can be solved according to various rules parametrized by $\alpha$, e.g. $\alpha = 0$ (Ito), 0.5 (Stratonovich) and 1, with different properties. This leads, in particular, to different predictions for the equilibrium distributions $p(z)$ and drift fields $d(z) = \left\langle \frac{\Delta z (z)}{\Delta t} \right\rangle$ depending on $\alpha$ \cite{Brettschneider2011}. We can compare such predictions with our measurements in order to determine the correct value of $\alpha$ (which turns out to be $\alpha = 1$). We remark that this is only possible because we have \emph{a priori} knowledge of the force $F(z)$ acting on the particle, which in particular amounts to gravity (buoyancy) for $z>280\,\mathrm{nm}$ [Fig. 1(a)], a fact overlooked by Mannella and McClintock in their Comment \cite{Comment} because not clearly stated in our Letter \cite{Volpe2010}. The details follow.

$F(z)$ has two components: gravity (buoyancy) $-G_{\mathrm{eff}} = -5.9\,\mathrm{fN}$, which pushes the particle towards the wall and is constant for all $z$, and electrostatic interactions $F_{\mathrm{el}} = Be^{-\kappa z}$, which prevents the particle from sticking to the wall and decays away from the wall with $\kappa^{-1} = 18\,\mathrm{nm}$. For $z>280\,\mathrm{nm}$, in particular, there is only effective gravity, which is known without any fitting parameter [black line in Fig. 1(a)]. We remark that these forces act on the particle independently of the presence of Brownian noise, i.e. they would also be present if the Brownian noise were switched off, for example, by decreasing the temperature of the system towards $T = 0$. The corresponding SDE is 
\begin{equation}
dz = -\frac{F(z)}{\gamma(z)} dt + \sqrt{2D(z)}dW,
\end{equation}
where $D(z)$ is the position-dependent diffusion coefficient $D(z)$, $\gamma(z)$ the particle friction coefficient and $W$ a Wiener process. The corresponding $\alpha$-dependent $p(z)$ and $d(z)$ can be related to force measurements as explained \cite{Volpe2010,Brettschneider2011}. In particular, the values of 
\begin{equation}
F_p(z) = \frac{k_B T}{p(z)} \frac{dp(z)}{dz}
\end{equation}
(solid red line) and 
\begin{equation}
F_d(z) = \gamma(z) d(z)
\end{equation}
(dashed blue line), which have units of force, are shown for $\alpha = 0$ in Fig. 1(b), $\alpha = 0.5$ in Fig. 1(c) and for $\alpha = 1$ in Fig. 1(d). We remark that, even though $F_p(z)$ and $F_d(z)$ are clearly different, $F_d(z) - F_p(z)$ is independent from $\alpha$, as correctly pointed out by Mannella and McClintock \cite{Comment}. 

In Ref. \cite{Volpe2010}, we experimentally measured $F_p(z)$ (red squares) and $F_d(z)$ (blue circles) [Figs. 1(b-e)]: there is agreement with the solution of the SDE (1) for $\alpha = 1$ [Fig. 1(f-g)], while the cases of $\alpha$ = 0 and 0.5 show clear deviations [Figs. 1(b-e)]. This has the consequences that the force $F(z) = F_p(z)$ and $F(z) = F_d(z) - \gamma(z) \frac{dD(z)}{dz}$, which is the main result of Refs. \cite{Volpe2010} and \cite{Brettschneider2011}. Finally, we remark that for other systems, which are not coupled to a heat-bath \cite{nonthermal}, the relations between $F(z)$, $F_p(z)$ and $F_d(z)$ may be different.

$ $\\
Giovanni Volpe$^{1,2}$, Laurent Helden$^{2}$, Thomas Brettschneider$^{2}$, Jan Wehr$^{3}$, and Clemens Bechinger$^{1,2}$\\
$^1$Max-Planck-Institut f\"{u}r Intelligente Systeme, Heisenbergstra{\ss}e 3, 70569 Stuttgart, Germany\\
$^2$2. Physikalisches Institut, Universit\"{a}t Stuttgart, Pfaffenwaldring 57, 70550 Stuttgart, Germany\\
$^3$Department of Mathematics, University of Arizona, Tucson, AZ 85721-0089, USA



\begin{thebibliography}{10}

\bibitem{Volpe2010} G. Volpe \emph{et al.}, Phys. Rev. Lett. \textbf{104}, 170602 (2010).

\bibitem{Brettschneider2011} T. Brettschneider \emph{et al.}, Phys. Rev. E \textbf{83}, 041113 (2011).

\bibitem{Comment} R. Mannella and P. V. E. McClintock, Comment on "Influence of Noise on Force Measurements.лл


\bibitem{nonthermal} Some examples of such systems are given, e.g., in R. Kupferman \emph{et al.}, Phys. Rev. E \textbf{70}, 036120 (2004) and P. Ao \emph{et al.}, Complexity \textbf{12}, 19-27 (2007) and references therein.


\end{thebibliography}
\end{document}